\documentclass[aps,prd,reprint,groupedaddress,nofootinbib]{revtex4-1}

\usepackage[utf8]{inputenc}
\usepackage{mathrsfs }
\usepackage{mathtools}
\usepackage{amssymb}
\usepackage{booktabs}
\usepackage{braket,slashed,bm}
\usepackage{array,multirow}
\usepackage{xparse}
\usepackage{etoolbox}
\usepackage{multirow}
\usepackage{footmisc}
\usepackage{color}
\usepackage{multirow}
\usepackage{xspace}
\usepackage{listings}
\usepackage{changepage}

\usepackage{slashed}
\usepackage{xcolor,colortbl}
\usepackage{hyperref}

\usepackage{comment}
\usepackage{relsize}
\usepackage{subfig}
\usepackage{float}
\usepackage{geometry}
\usepackage{float}
\usepackage{braket,slashed,bm}
\usepackage{array,multirow}
\usepackage[normalem]{ulem}
\usepackage{xcolor,cancel,youngtab}
\usepackage{mathrsfs}
\usepackage{booktabs}
\usepackage{adjustbox}
\usepackage{mathtools}
\usepackage{soul}
\usepackage{xparse}
\usepackage{etoolbox}
\usepackage{hypcap}
\usepackage{amsmath,amsfonts,amssymb,stackengine,graphicx}
\usepackage[utf8]{inputenc}
\usepackage{graphicx}
\def\ltap{\raisebox{-.4ex}{\rlap{$\sim$}} \raisebox{.4ex}{$<$}} 
\def\gtap{\raisebox{-.4ex}{\rlap{$\sim$}} \raisebox{.4ex}{$>$}}
\def\beq{\begin{equation}}
\def\eeq{\end{equation}}
\def\barr{\begin{array}}
\def\earr{\end{array}}
\def\dis{\displaystyle}

\definecolor{darkcyan}{cmyk}{1,0,0,0.4}

\newcommand{\be}{\begin{equation}}
\newcommand{\ee}{\end{equation}}
\newcommand{\bea}{\begin{eqnarray}}
\newcommand{\eea}{\end{eqnarray}}

\newcommand{\bi}{\begin{itemize}}
	\newcommand{\ei}{\end{itemize}}

\begin{document}
\title{Neutron oscillation and Baryogenesis from six dimensions}
\author{Mathew Thomas Arun$^1$, 
Debajyoti Choudhury$^2$}
\affiliation{$^1$School of Physics, IISER Thiruvananthapuram, Vithura, Kerala, 695551, India\\
$^2$Department of Physics and Astrophysics, University of Delhi, Delhi 110005, India}
\date{January 2022}

\begin{abstract}
Considering a six-dimensional geometry orbifolded on $S^1/Z_2\times
S^1/Z_2$ with quarks and leptons localised on orthogonal branes, we
show that the construction admits observable $n-\bar{n}$ oscillation
while naturally suppressing the proton decay rates. Consistent with
other low-energy observables, the model also accommodates baryogenesis
at $\mathcal{O}$(10 TeV) scale.
\end{abstract}
\maketitle

Baryon number ($B$) violation, a key ingredient for generating the
observed baryon to photon ratio in the universe ($n_B/n_\gamma =
(6.19\pm 0.14)\times 10^{-10}$), is allowed in the Standard Model (SM)
only through non-perturbative
processes~\cite{tHooft:1976rip,*tHooft:1976snw,*Kuzmin:1985mm,*Dolgov:1991fr}. Presumably
important for baryogenesis at temperatures near the electro-weak
scale~\cite{Shaposhnikov:1987tw,*Babu:2006xc,*Babu:2006wz,*Babu:2008rq,*Morrissey:2012db,*Dev:2015uca},
such processes are highly suppressed at low temperatures, and $B$ is
seemingly a good global symmetry. With $\cancel B$ and CP violation
within the SM not being large enough to generate the $n_B/n_\gamma$,
other sources must be explored.

The simplest gauge invariant and $\cancel B$ effective operator also
violates lepton number ($L$) by one unit, and non-observation of
proton decay pushes the corresponding scale to well above $10^{15}$
GeV. On the other hand, if the leading operator were a $\Delta B = 2$
one, the scale could be much lower and be probed by looking for either
$n-\bar{n}$ oscillation in nuclei, or annihilations brought about in
collisions of cold neutrons against a target. The characteristic
time scales are related through $T_{\rm nucl} =
\tau^2_{\rm free} R $, where $R$ characterises the strong interaction
$\bar{n}$ annihilation time, and is, typically, ${\cal O}(100\,{\rm
  MeV})$~\cite{Friedman:2008es}.  Using different nuclei, experiments
at SOUDAN-II (Fe$_{56}$)~\cite{Chung:2002fx}, Super-Kamiokande
(O$_{16}$)~\cite{Super-Kamiokande:2011idx} and SNO
(deuteron)~\cite{Bellerive:2016byv} have constrained $\tau_{\rm free}$ to
be larger than $1.3 \times 10^8$s, $2.7 \times 10^8$s and $1.23 \times
10^8$s respectively, each at 90\% C.L.  Relating to the underlying
theory is best done\cite{Mohapatra:1980qe,*Ozer:1982qh,*Kuo:1980ew} by
considering the matrix element of the effective six-quark
(dimension-9) operator {\em viz.}  $ \Delta m \equiv \tau_{\rm free}^{-1}
= \langle \bar{n} | \mathcal{O}_9 | n \rangle$. The transition
probability for pure state $| n\, ; t= 0 \rangle$ 
to evolve to 
$|\bar{n} \, ; t\rangle$ is given by
$P(t) = (t/\tau_{\rm free})^2 e^{-\lambda t}$
where $\lambda^{-1} = 880 \, s$ is the mean life of a free neutron. The
bound on $\tau_{\rm free}$ implies $\Delta m \lesssim 6 \times
10^{-33}$ GeV, or, for an ${\cal O}(1)$ Wilson coefficient, a new
physics scale ($\gtrsim 500$ TeV~\cite{Arnold:2012sd})
much lower
than the proton decay scale.
A roadblock to a UV-complete model for
$n-\bar{n}$ oscillation is that it, generically, needs two new fields with gauge
symmetry allowing one of these to couple to a $\Delta L = 1$ current
as well, thereby requiring an unnatural suppression for the said coupling.

In our quest for a well-motivated scenario that 
naturally circumvents all such
constraints, we propose a six-dimensional
space-time orbifolded on $S^1/Z_2\times
S^1/Z_2$~\cite{Choudhury:2006nj,Arun:2016csq,Arun:2017zap} and a
highly-warped $x_5-$direction. With quarks and leptons localised on
orthogonal branes, the geometry supports substantial $\Delta
\mathcal{B} = 2$ while evading proton-decay constraints {\em without}
any hierarchy/un-naturalness in the parameters.

With successive warpings along the two compactified
dimensions, ($x_{4} \in [0,\pi R_y]$ and $x_{5} \in [0,\pi r_z]$) that
are individually $Z_2$-orbifolded with 4-branes sitting at each of the
edges, the geometry is described by the line
element~\cite{Choudhury:2006nj}
\[
ds^2_6= b^2(x_5)[a^2(x_4)\eta_{\mu\nu}dx^{\mu}dx^{\nu}+dx_4^2]+dx_5^2 \ ,
\]
where $\eta_{\mu \nu}$ is the flat metric. 
Denoting the fundamental scale in six dimensions by $M_6$ and the negative bulk cosmological constant by 
$\Lambda_6$, 
the total bulk-brane Lagrangian is, thus,
\[
\barr{rcl}
{\cal L} &=& \dis \sqrt{-g_6} \, 
   (M_{6}^4R_6-\Lambda_6)\\[1ex]
&+ & \dis\sqrt{-g_5}\, 
      [V_1(x_5) \, \delta(x_4)+V_2(x_5) \, \delta(x_4-\pi R_y)]\\[1ex]
&+& \dis \sqrt{-\tilde g_5} \, 
     [V_3(x_4) \, \delta(x_5)+V_4(x_4) \, \delta(x_5-\pi r_z)] \, .
\earr
\]
The five-dimensional metrics ($g_5, \tilde g_5$) are those
induced on the appropriate 4-branes, and the brane potentials $V_i$
encode the Israel junction conditions.

Generalizing from the restrictive case studied in
Ref.\cite{Choudhury:2006nj}, we  admit a five-dimensional induced non-zero cosmological
constant $\widetilde{\Omega} < 0$ on the 4-branes, allowing these
to be bent.  The four-dimensional cosmological constant, though, is held
 to be zero (the generic
case~\cite{Arun:2016csq} only leads to algebraic complications). With
this, the Einstein equations lead to
\beq
\label{generic_solz}
b(x_5) = b_1 \cosh(k |x_5| + b_2) \ , \, 
b_1 = {\rm sech}(k \pi r_z + b_2) \ ,
\qquad 
\eeq
with $k = \sqrt{-\Lambda_6/10 \, M_6^4}$. Similarly,
\beq
a(x_4) = \exp(- c |x_4|) \, , 
\text{where} \ \ c \equiv k b_1 R_y / r_z\, .
  \eeq
  Unlike
  ref.\cite{Choudhury:2006nj}, 
we consider a very large $b_2$
\cite{Arun:2016csq,Arun:2017zap} leading to $b(x_5) \propto 
\exp(k |x_5|)$, and one is forced to $c \ll k$, unless a large,
and unpleasant, hierarchy between $R_y$ and $r_z$ is to be
admitted. This also implies that the 4-brane tensions are 
pairwise almost equal
and opposite (
$V_1 = - V_2  \approx  \dis  0$ and 
$V_3 = - V_4  \approx  \dis - 8 M^4 k$) 
thereby ensuring 
the near vanishing of the
induced cosmological constant on the 4-branes at the ends of the world.
In this limit, the metric is nearly conformally flat,
and, along with the $AdS_6$ bulk,  resembles a
generalization of the Randall-Sundrum geometry to one dimension higher.

In this space-time, the Lagrangian for a free massless
  fermion is given by
\[
\barr{rcl}
  {\cal L}_{\rm fermion} &=& \dis \frac{1}{2} b^4(x_5)\,
  \Big(\bar{Q} \, i \Gamma^{a} \, E^{M}_{a} \, D_M \, Q\ \Big)
  \overline\Delta(x_4) 
\\[1ex]
&+ & \dis  b^5(x_5)\, \Big(\bar{L} \, i\Gamma^{a} \, E^{M}_{a}
\, D_M \, L\ \Big)\delta(x_5)
\\[1ex]
\dis \overline\Delta(x_4) & \equiv & \dis \delta(x_4)+\delta(x_4-\pi R_y)
\earr
\]
where $Q \, (L)$ denote the quark (lepton) fields and $E^M_a$ are the
appropriate f\"unfbeins. The low-energy
phenomenology is, of course, dictated by only
the zero-modes of the fermions. Since the $x_4-$direction is nearly
flat, so would be the wave profile of the leptonic zero mode. On the
other hand, the quark zero mode is given, to a very good
approximation, by an exponential function.


To induce $\cancel{\cal B}$, we begin by
introducing two colored scalar fields $\phi (3,1,-1/3)$ and $\omega
(6,1,2/3)$, localised, of course, on the two 4-branes at $x_4= 0$ and
$x_4= \pi R_y$.  With leptons being confined to the 4-brane at $x_5 =
0$, the scalar interactions are given by
\begin{eqnarray}
  \mathcal{L}_{\rm scal} &=& \dis 
  \sqrt{2r_z}\Big[ y_{ud} \phi \overline{u^c} d +
  \sqrt{4 r_z R_y} y_{ue} \phi^* \overline{u^c} e \ \delta(x_5)  \nonumber \\
  &+& \dis z_{dd} \omega \overline{d^c} d 
  +  \lambda M \phi^2 \omega \Big]\overline \Delta(x_4) + h.c.\, .
\end{eqnarray}
All fermions here are $SU(2)$ singlets and we have included just the
relevant term of the potential. Introduced for convenience, the scale
$M$ is equated to the mass of the heaviest scalar field.  Since, post
compactification, we would be interested only in the lightest
KK-modes, the five dimensional fields (${\cal F} = u, d, \phi,
\omega$) could be decomposed as ${\cal F}(x_\mu, x_5) = (2 r_z)^{-1/2}
      {\cal F}(x_\mu)\, \chi_{\cal F}(x_5)$, where the zero-mode
      wavefunctions $\chi_{\cal F}(x_5)$ satisfy the normalization
      condition $\int dx_5 b^s \chi^2_{\cal F} = 1$, with $(s=2,3)$
      for scalars and fermions respectively, yielding canonically
      normalized four-dimensional fields.  In contrast, thanks to
      the $x_4$-direction being nearly flat, the wavefunction for
      the lepton zero-mode is a trivial one. To obtain the effective
      four-dimensional theory, one needs to integrate over both
      $x_5$ and $x_4$ (given the smallness of the warping $c$, the
      second integration is essentially trivial), resulting in
\begin{equation}
  \barr{rcl}
  \mathcal{L}_{int} &=& \dis \eta_{ud} \phi \overline{u^c} d
  + \eta_{ue} \phi^* \overline{u^c} e
  +  \zeta_{dd} \omega^* \overline{d^c} d
  \\
  & + & \dis \rho M  \phi^2 \omega + h.c. \ ,
  \earr
\label{eq:4dinteraction}
\end{equation}
where
\begin{equation}
\barr{rcl}
\zeta_{dd} & = & \dis (2 r_z)^{-1} z_{dd} \int dx_5 b^2 \chi_\phi \chi_d^2
\\
\eta_{ud} & = &  \dis (2 r_z)^{-1} y_{ud} \int dx_5 b^2 \chi_\phi \chi_u \chi_d
\\
\rho & = &  \dis (2 r_z)^{-1}\lambda \int dx_5 b^2 \chi^2_\phi \chi_\omega
\\
\eta_{ue} & = & \dis y_{ue} b^2(0) \chi_\phi(0) \chi_u(0)
\earr
\label{eqn:coup_integral}
\end{equation} 
The relations above may include any possible flavour structure, a
point that we return to later. Now, for fermions with vanishing bulk
masses (an excellent approximation for the first and second generation
fields), the wave profiles would be expected to be nearly identical,
and the ratio $\zeta_{dd} / \eta_{ud} \approx z_{dd}/y_{ud}$.

The Wilson coefficients for proton-decay and neutron oscillation are
now straightforward, viz.
\begin{equation}
  C^{p} = \frac{\eta_{ud} \eta_{ue}}{ m_{\phi}^2}  \ , \quad 
   C^{nn} = \frac{\rho M}{m_{\phi}^2 m_\omega^4} \eta_{ud}^2 \, \zeta_{dd} \, .
 \label{pdecay_nn}
\end{equation}
Qualitative features of this result are best appreciated by assuming
that all the masses are of the same order. Despite the fact that
$C^{nn}$ is fourth order in the couplings while $C^p$ is only
bilinear, $C^p \ll C^{nn}$ on account
of the factor $b^2(0)$ in $\eta_{ue}$. Note that this relative
suppression does not need any hierarchy,
either of couplings or
between masses.  Rather, it is engendered dynamically on account of
the warping.

The neutron oscillation rate is determined by
$|\Delta m| = |\langle\bar{n}|H_{eff}|n\rangle| = 8 \xi^2 C^{nn}/3$
where the matrix element is computed using vacuum insertion
approximation~\cite{Arnold:2012sd} with $\xi$ parametrizing the
reduced matrix element for the three quark operator. Lattice
computations~\cite{CP-PACS:2004wqk} give $\xi \approx 0.0096
{}^{+6}_{-20}\, {\rm
GeV}^3$, with this error being the main
uncertainty in the calculation.

The observed limits on $|\Delta m|$ can now be translated to
constraints on the parameter space. In view of the large
dimensionality of the latter, it is instructive to make the
simplifying assumption that $\eta_{ud} = \zeta_{dd}
= \rho \equiv \kappa$ and $m_\phi = m_\omega = M$. The saturation
of the limit can then be expressed as
\begin{equation}
\frac{|\Delta m|}{|\Delta m|_{\rm max}}
     \approx \left[\frac{\kappa}{0.01}\right]^4 \,
             \left[\frac{M}{15 \, {\rm TeV}}\right]^{-5} \, .
\end{equation}
More precise values are listed in Table~\ref{tab:const} for a few
benchmark points. It is worth noting here that, for $k\,r_z \sim 8.5$
(a value that that leads to $C^p \ltap 10^{-30}$, thereby suppressing
proton decay to rates well below the current limit) also implies that
for anarchic bulk-couplings $z_{dd},
y_{ud}, \lambda \sim \mathcal{O}(1)$, the corresponding 4-dimensional
ones ($ \zeta_{dd}, \eta_{ud}, \rho$), obtained on integrating over
$x_5$, are all $ \sim {\cal O}(10^{-2})$.  Further suppression of the
effective couplings could arise in two ways: $(i)$ starting with
smaller bulk couplings or $(ii)$ localising the light quark states
away from the scalars. In contrast, the scenario with $\kappa\sim 1 $
is achievable only for a flat extra dimension, at the cost of very
large scalar masses and self-consistency of the treatment would imply
that the scales of the KK-excitations be at least as large, thereby
taking the entire scenario beyond detection.
\begin{table}[!hb]
\begin{center}
\begin{tabular}{|c|c|c|c|c|}
\hline 
$\kappa$ & 1 & 0.1 & 0.01 & 0.001 \\ \hline
$m_{\phi} ({\rm TeV}) = m_\omega = M$ & 670 & 106 & 17 & 2.6 \\ \hline
$m_{\phi}({\rm TeV})=m_\omega/3=M/3 $  & 345 & 55 & 9 & 1.4 \\ \hline
\end{tabular}
\caption{Lower limits on the mass parameter $m_\phi$ as a function of the
common coupling.}
\label{tab:const}
\end{center}
\end{table}

We now examine other possible phenomenological consequences. For light
(a few TeVs) $\phi$ or $\omega$, resonance production at the LHC is
possible, leading to a peak in the dijet invariant mass
(while a lepton-jet decay is notionally available to the $\phi$, it is
highly suppressed). With the QCD background being very large, the only
hope is to concentrate on high-$p_T$, high invariant-mass events. For
$m_\phi = 3 TeV$ and $\eta_{ud} =
10^{-2}$, the $\phi$-production cross-section is $\lesssim 0.4$fb (and somewhat lower for the
$\omega$) even without accounting for efficiencies. In other words,
even in the most optimistic scenario, direct detection would have to
wait. The only caveat to this would be to consider a hierarchy in the
scalar masses whereby one of them could be made significantly lighter
and brought into the current reach of the LHC.

Much more important are the flavour sector observables. While
$n$-$\bar n$ oscillation needs only the $d-d-\omega$ coupling, it is
conceivable that the $s-s-\omega$ coupling is unsupressed as
well. This would result in an effective flavour changing Hamiltonian
of the form ${\cal H}^{\Delta S = 2} = m_\phi^{-2} (\bar{s^c} P_R
s)(\bar d P_L d^c)$, thereby contributing to $K^0$--$\bar{K^0}$
oscillation (which, within, the SM, proceeds through the
charm-dominated box diagram). To bring this to the usual form, one has
to effect Fierz rotations~\cite{Chakraverty:2000df}, both in the Dirac
space and the color space, using $(6_c \otimes 6_c)_1 = (2/3) (1_c
\otimes 1_c) + (1/2) (8_c \otimes 8_c)_1$, encapsulating both
color-unsuppressed and suppressed contributions. Comparing with the SM
contribution (which saturates the observed mass difference), we have
\begin{equation}
    \frac{Re(\zeta_{dd} \zeta_{ss}^*)}{8 m_{\omega}^2} \ltap
    \frac{G_F^2 M_W^2}{16 \pi^2}\mathcal{F}^0 \sim 10^{-7} {\rm TeV}^{-2}\, ,
\end{equation}
where $\mathcal{F}^0$ includes the box-diagram
computation~\cite{FlavourLatticeAveragingGroup:2019iem} (note that the
hadronic matrix element calculation is common to both). Assuming that
the couplings are of similar size, for $m_{\omega} = 3\,$TeV, this
translates to $|\zeta_{dd}| < 0.0026$.
In other words, the coupling values required for 
observable neutron-oscillation rates more than easily satisfy
the bounds from kaon oscillations. This exercise, though, tells us that any
introduction of hierarchy in the scalar masses needs to be approached
carefully.

As is well-known, the creation of a baryon asymmetry needs not only
$\cancel B$ and $\cancel {CP}$ (much larger than that we have in the
SM) but also an accompanying epoch with out-of-equilibrium
condition. To create an environment amenable for this, let us
augment the model with the inclusion of a further copy of $\phi$
(henceforth called $\phi_1$) and a singlet $S$ (which, for
simplicity, we consider to be localized on the same branes as the
colored scalars). Similarly, we also assume, that the $\phi$ mass
matrix is diagonal and that there is no substantial mass hierarchy
between these.  As for the potential in \ref{eq:4dinteraction}, it
now includes additional terms such as ($i=1,2$)
\begin{equation}
  \barr{rcl}
  -V &\ni& \dis 
  M \Big[\rho_{ij} \phi_i \phi_j  \omega +
       \widetilde \rho_{ij} \phi_i^* \phi_j S 
       \Big] + h.c. \, ,
\earr
  \label{eq:extra_terms_in_pot}
\end{equation}
where, for the sake of simplicity, we
retain only the zero-modes.  While the terms in
eq.(\ref{eq:extra_terms_in_pot}) could have extra factors,
of $S/M$, these do not add anything to our discussion,
and we omit them.

At the tree level, we forbid any Yukawa couplings
for the new field $\phi_2$ (this could be arranged simply by
introducing a softly broken discrete symmetry). Thus, the dominant
decay available to it would be one through $\rho_{12}$ leading to
$\phi_2 \to u d d d$, or a final state of $B = 4/3$. On the other
hand, $\phi_1 \to \bar u \bar d$ ($B = -2/3$).

\begin{figure}[!t]
\vspace {-1.5cm}
    \includegraphics[width=8cm,height=7cm]{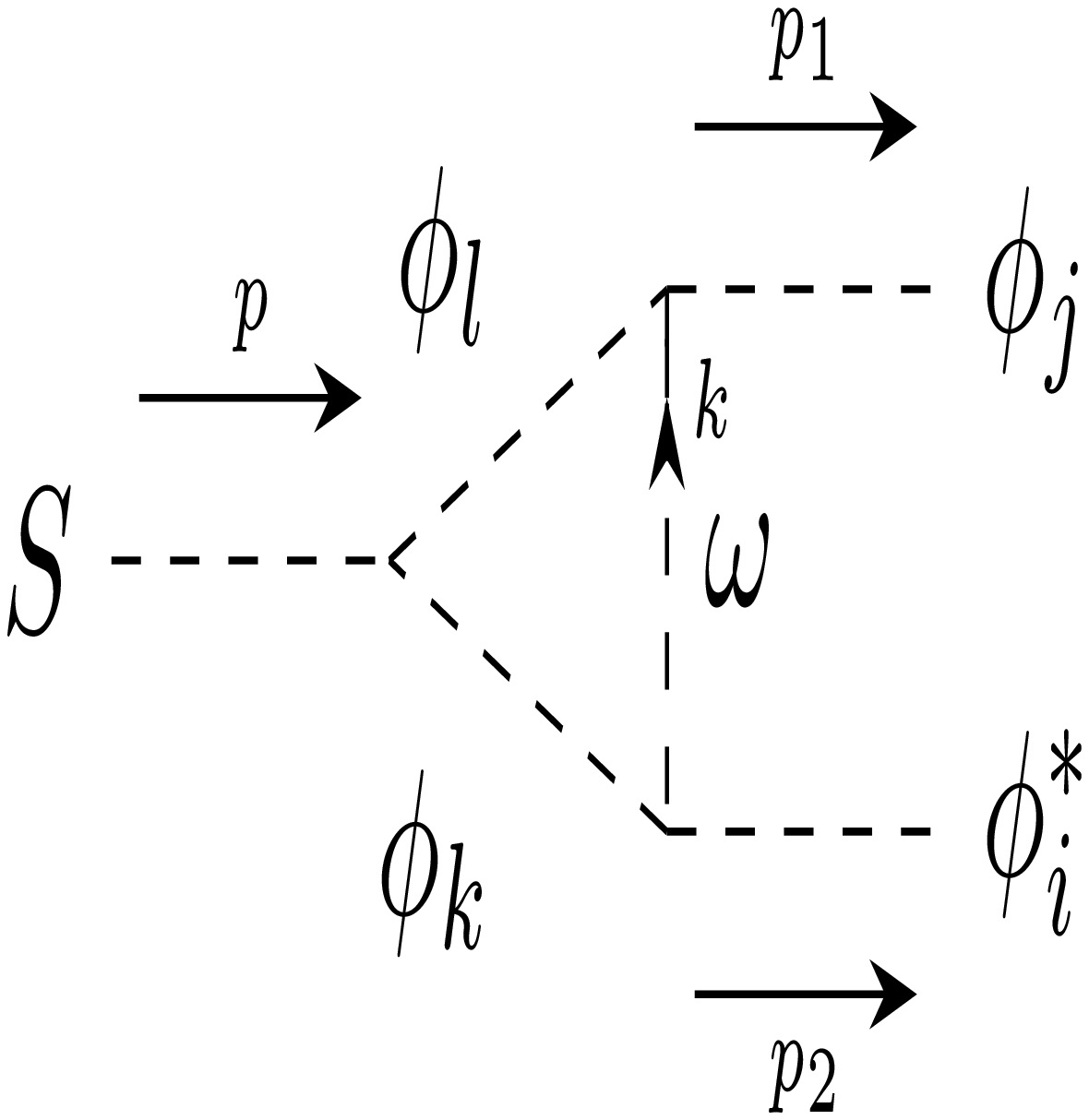}
\vspace {-3.5cm}
\caption{One-loop diagram contributing to $B$-asymmetry generated from
a  decaying $S$. }
\label{baryo}
\end{figure}

For $2 m_\phi <
m_S \ltap 2m_\omega$, the dominant decay of the $S$ would be
$S \to \phi_i^* \phi_j$. While the tree-level diagram is obvious,
one-loop corrections are brought about by the diagram in 
Fig.\ref{baryo}.
The latter evidently has a non-zero absorptive part, receiving
contributions from possibly three ways of cutting a pair of internal
lines. For example, for $m_{\phi_1} \sim m_{\phi_2} \gtap m_\omega /
2$, only the vertical cut in Fig.~\ref{baryo} contributes.  
Since a net baryon number $\epsilon_B$ (per $S$-decay)
can arise only from the difference
$\Gamma(S \to \phi_1 \phi_2^*) - \Gamma(S \to \phi_1^* \phi_2)$,
we have
(denoting $x_a \equiv m_a^2 / m_S^2$) 
\begin{eqnarray}
  \epsilon_B &\approx &\dis \frac{M^2}{4\pi^2 m_{S}^2 \beta} \, 
\log\Big(\frac{x_\omega-x_\phi + 1+\beta}{x_\omega-x_\phi+1-\beta}\Big) \; {\cal A}_\rho
  \nonumber \\
{\cal A}_\rho  &\equiv & \dis
Im\left[\widetilde\rho_{12}^* (\rho^\dagger \widetilde\rho \rho)_{12}\right]/
\sum_{i,j} |\widetilde\rho_{ij}|^2
\label{eq:asymmetry}
\end{eqnarray}
where $\beta = (1-4 x_{\phi})^{1/2}$ and we have neglected terms
  proportional to the mass difference $\delta m_\phi^2$.

Thus, the number density $n_S$ of the decaying field is related
to those for the (anti-)baryons through,
$n_b - n_{\bar{b}} = n_S \epsilon_B B_q$, where $B_q=2$ is
the baryon number created per decay. 
Since the photons in the universe far outnumber the
baryons, they dominate the entropy density $s$ of the universe and we may
write~\cite{Cline:2006ts}
\begin{equation}
  \frac{n_b - n_{\bar{b}}}{s} = \frac{n_S B_q \epsilon_B}{4 \pi^2/45 g_* T^3}
  = \frac{180 \zeta(3)}{\pi^4} \epsilon_B \ ,
\label{eq:nns}
\end{equation}
where $g_* \approx 107$ is the weighted number of degrees of freedom
operative at that temperature. 

 Using Eq.\ref{eq:nns}, the constraint on baryon asymmetry in
 the universe as, $\epsilon_B = 3.8 \times 10^{-11}$.
This could be arranged quite easily in the model. In the parameter
space where the nuetron-anti-neutron oscillation is observable, 
$m_\phi \sim m_\omega = 3 TeV $, and  $M\sim m_S = 10 TeV$, the
asymmetry required for baryogenesis is obtained for
${\cal A}_\rho \lesssim 2.2 \times 10^{-9}$.
Note that the very structure
of ${\cal A}_\rho$ implies cancellations between the nominal phases in
the couplings.  Consequently, the residual phase is naturally
smaller. Indeed, such a phase is also stipulated by the $\epsilon'$
parameter in kaon oscillations.

To summarise, we have investigated baryon-number violation in a
six-dimensional world compactified on $(S^1/Z_2) \otimes (S^1/Z_2)$
wherein one direction is highly warped with the other being almost
flat (minuscule warping).  Boasting stabilised moduli~\cite{Arun:2016csq},
such models are more natural than their
flat extra dimension counterparts.  As we have demonstrated here,
augmenting the minimal SM extension by just two
extra colored scalars allows for low-energy $\Delta B = 2$ processes,
thereby offering the tantalizing prospect of an observable $n-\bar{n}$
oscillation signal while proton decay is dynamically suppressed
(without the need for any hierarchy). While, for natural sizes of
Yukawa couplings, the masses of the scalars are just beyond the reach
of the LHC, these are likely to be visible in the next generation of
hadronic colliders or their presence inferred from more sensitive
flavour probes. Most interestingly, a minor addition to the spectrum
opens the possibility for multi-TeV scale baryogenesis. The parameter
space amenable to reproducing the observed baryon asymmetry is large
and no hierarchy or unnatural assumptions need be
invoked. Furthermore, these scalars hold the prospect of
generating strong phase transitions, signals of which are likely to be
observable in the next generation of gravitational wave detectors.

{\em M.T.A. acknowledges financial support of the DST through INSPIRE Faculty grant [DST/INSPIRE/04/2019/002507]}. 

\bibliographystyle{apsrev4-1}
\bibliography{bibliography.bib}

\begin{thebibliography}{25}%
\makeatletter
\providecommand \@ifxundefined [1]{%
 \@ifx{#1\undefined}
}%
\providecommand \@ifnum [1]{%
 \ifnum #1\expandafter \@firstoftwo
 \else \expandafter \@secondoftwo
 \fi
}%
\providecommand \@ifx [1]{%
 \ifx #1\expandafter \@firstoftwo
 \else \expandafter \@secondoftwo
 \fi
}%
\providecommand \natexlab [1]{#1}%
\providecommand \enquote  [1]{``#1''}%
\providecommand \bibnamefont  [1]{#1}%
\providecommand \bibfnamefont [1]{#1}%
\providecommand \citenamefont [1]{#1}%
\providecommand \href@noop [0]{\@secondoftwo}%
\providecommand \href [0]{\begingroup \@sanitize@url \@href}%
\providecommand \@href[1]{\@@startlink{#1}\@@href}%
\providecommand \@@href[1]{\endgroup#1\@@endlink}%
\providecommand \@sanitize@url [0]{\catcode `\\12\catcode `\$12\catcode
  `\&12\catcode `\#12\catcode `\^12\catcode `\_12\catcode `\%12\relax}%
\providecommand \@@startlink[1]{}%
\providecommand \@@endlink[0]{}%
\providecommand \url  [0]{\begingroup\@sanitize@url \@url }%
\providecommand \@url [1]{\endgroup\@href {#1}{\urlprefix }}%
\providecommand \urlprefix  [0]{URL }%
\providecommand \Eprint [0]{\href }%
\providecommand \doibase [0]{http://dx.doi.org/}%
\providecommand \selectlanguage [0]{\@gobble}%
\providecommand \bibinfo  [0]{\@secondoftwo}%
\providecommand \bibfield  [0]{\@secondoftwo}%
\providecommand \translation [1]{[#1]}%
\providecommand \BibitemOpen [0]{}%
\providecommand \bibitemStop [0]{}%
\providecommand \bibitemNoStop [0]{.\EOS\space}%
\providecommand \EOS [0]{\spacefactor3000\relax}%
\providecommand \BibitemShut  [1]{\csname bibitem#1\endcsname}%
\let\auto@bib@innerbib\@empty
\bibitem [{\citenamefont {'t~Hooft}(1976{\natexlab{a}})}]{tHooft:1976rip}%
  \BibitemOpen
  \bibfield  {author} {\bibinfo {author} {\bibfnamefont {G.}~\bibnamefont
  {'t~Hooft}},\ }\href {\doibase 10.1103/PhysRevLett.37.8} {\bibfield
  {journal} {\bibinfo  {journal} {Phys. Rev. Lett.}\ }\textbf {\bibinfo
  {volume} {37}},\ \bibinfo {pages} {8} (\bibinfo {year}
  {1976}{\natexlab{a}})}\BibitemShut {NoStop}%
\bibitem [{\citenamefont {'t~Hooft}(1976{\natexlab{b}})}]{tHooft:1976snw}%
  \BibitemOpen
  \bibfield  {author} {\bibinfo {author} {\bibfnamefont {G.}~\bibnamefont
  {'t~Hooft}},\ }\href {\doibase 10.1103/PhysRevD.14.3432} {\bibfield
  {journal} {\bibinfo  {journal} {Phys. Rev. D}\ }\textbf {\bibinfo {volume}
  {14}},\ \bibinfo {pages} {3432} (\bibinfo {year} {1976}{\natexlab{b}})},\
  \bibinfo {note} {[Erratum: Phys.Rev.D 18, 2199 (1978)]}\BibitemShut {NoStop}%
\bibitem [{\citenamefont {Kuzmin}\ \emph {et~al.}(1985)\citenamefont {Kuzmin},
  \citenamefont {Rubakov},\ and\ \citenamefont {Shaposhnikov}}]{Kuzmin:1985mm}%
  \BibitemOpen
  \bibfield  {author} {\bibinfo {author} {\bibfnamefont {V.~A.}\ \bibnamefont
  {Kuzmin}}, \bibinfo {author} {\bibfnamefont {V.~A.}\ \bibnamefont {Rubakov}},
  \ and\ \bibinfo {author} {\bibfnamefont {M.~E.}\ \bibnamefont
  {Shaposhnikov}},\ }\href {\doibase 10.1016/0370-2693(85)91028-7} {\bibfield
  {journal} {\bibinfo  {journal} {Phys. Lett. B}\ }\textbf {\bibinfo {volume}
  {155}},\ \bibinfo {pages} {36} (\bibinfo {year} {1985})}\BibitemShut
  {NoStop}%
\bibitem [{\citenamefont {Dolgov}(1992)}]{Dolgov:1991fr}%
  \BibitemOpen
  \bibfield  {author} {\bibinfo {author} {\bibfnamefont {A.~D.}\ \bibnamefont
  {Dolgov}},\ }\href {\doibase 10.1016/0370-1573(92)90107-B} {\bibfield
  {journal} {\bibinfo  {journal} {Phys. Rept.}\ }\textbf {\bibinfo {volume}
  {222}},\ \bibinfo {pages} {309} (\bibinfo {year} {1992})}\BibitemShut
  {NoStop}%
\bibitem [{\citenamefont {Shaposhnikov}(1987)}]{Shaposhnikov:1987tw}%
  \BibitemOpen
  \bibfield  {author} {\bibinfo {author} {\bibfnamefont {M.~E.}\ \bibnamefont
  {Shaposhnikov}},\ }\href {\doibase 10.1016/0550-3213(87)90127-1} {\bibfield
  {journal} {\bibinfo  {journal} {Nucl. Phys. B}\ }\textbf {\bibinfo {volume}
  {287}},\ \bibinfo {pages} {757} (\bibinfo {year} {1987})}\BibitemShut
  {NoStop}%
\bibitem [{\citenamefont {Babu}\ \emph {et~al.}(2006)\citenamefont {Babu},
  \citenamefont {Mohapatra},\ and\ \citenamefont {Nasri}}]{Babu:2006xc}%
  \BibitemOpen
  \bibfield  {author} {\bibinfo {author} {\bibfnamefont {K.~S.}\ \bibnamefont
  {Babu}}, \bibinfo {author} {\bibfnamefont {R.~N.}\ \bibnamefont {Mohapatra}},
  \ and\ \bibinfo {author} {\bibfnamefont {S.}~\bibnamefont {Nasri}},\ }\href
  {\doibase 10.1103/PhysRevLett.97.131301} {\bibfield  {journal} {\bibinfo
  {journal} {Phys. Rev. Lett.}\ }\textbf {\bibinfo {volume} {97}},\ \bibinfo
  {pages} {131301} (\bibinfo {year} {2006})},\ \Eprint
  {http://arxiv.org/abs/hep-ph/0606144} {arXiv:hep-ph/0606144} \BibitemShut
  {NoStop}%
\bibitem [{\citenamefont {Babu}\ \emph {et~al.}(2007)\citenamefont {Babu},
  \citenamefont {Mohapatra},\ and\ \citenamefont {Nasri}}]{Babu:2006wz}%
  \BibitemOpen
  \bibfield  {author} {\bibinfo {author} {\bibfnamefont {K.~S.}\ \bibnamefont
  {Babu}}, \bibinfo {author} {\bibfnamefont {R.~N.}\ \bibnamefont {Mohapatra}},
  \ and\ \bibinfo {author} {\bibfnamefont {S.}~\bibnamefont {Nasri}},\ }\href
  {\doibase 10.1103/PhysRevLett.98.161301} {\bibfield  {journal} {\bibinfo
  {journal} {Phys. Rev. Lett.}\ }\textbf {\bibinfo {volume} {98}},\ \bibinfo
  {pages} {161301} (\bibinfo {year} {2007})},\ \Eprint
  {http://arxiv.org/abs/hep-ph/0612357} {arXiv:hep-ph/0612357} \BibitemShut
  {NoStop}%
\bibitem [{\citenamefont {Babu}\ \emph {et~al.}(2009)\citenamefont {Babu},
  \citenamefont {Bhupal~Dev},\ and\ \citenamefont {Mohapatra}}]{Babu:2008rq}%
  \BibitemOpen
  \bibfield  {author} {\bibinfo {author} {\bibfnamefont {K.~S.}\ \bibnamefont
  {Babu}}, \bibinfo {author} {\bibfnamefont {P.~S.}\ \bibnamefont
  {Bhupal~Dev}}, \ and\ \bibinfo {author} {\bibfnamefont {R.~N.}\ \bibnamefont
  {Mohapatra}},\ }\href {\doibase 10.1103/PhysRevD.79.015017} {\bibfield
  {journal} {\bibinfo  {journal} {Phys. Rev. D}\ }\textbf {\bibinfo {volume}
  {79}},\ \bibinfo {pages} {015017} (\bibinfo {year} {2009})},\ \Eprint
  {http://arxiv.org/abs/0811.3411} {arXiv:0811.3411 [hep-ph]} \BibitemShut
  {NoStop}%
\bibitem [{\citenamefont {Morrissey}\ and\ \citenamefont
  {Ramsey-Musolf}(2012)}]{Morrissey:2012db}%
  \BibitemOpen
  \bibfield  {author} {\bibinfo {author} {\bibfnamefont {D.~E.}\ \bibnamefont
  {Morrissey}}\ and\ \bibinfo {author} {\bibfnamefont {M.~J.}\ \bibnamefont
  {Ramsey-Musolf}},\ }\href {\doibase 10.1088/1367-2630/14/12/125003}
  {\bibfield  {journal} {\bibinfo  {journal} {New J. Phys.}\ }\textbf {\bibinfo
  {volume} {14}},\ \bibinfo {pages} {125003} (\bibinfo {year} {2012})},\
  \Eprint {http://arxiv.org/abs/1206.2942} {arXiv:1206.2942 [hep-ph]}
  \BibitemShut {NoStop}%
\bibitem [{\citenamefont {Dev}\ and\ \citenamefont
  {Mohapatra}(2015)}]{Dev:2015uca}%
  \BibitemOpen
  \bibfield  {author} {\bibinfo {author} {\bibfnamefont {P.~S.~B.}\
  \bibnamefont {Dev}}\ and\ \bibinfo {author} {\bibfnamefont {R.~N.}\
  \bibnamefont {Mohapatra}},\ }\href {\doibase 10.1103/PhysRevD.92.016007}
  {\bibfield  {journal} {\bibinfo  {journal} {Phys. Rev. D}\ }\textbf {\bibinfo
  {volume} {92}},\ \bibinfo {pages} {016007} (\bibinfo {year} {2015})},\
  \Eprint {http://arxiv.org/abs/1504.07196} {arXiv:1504.07196 [hep-ph]}
  \BibitemShut {NoStop}%
\bibitem [{\citenamefont {Friedman}\ and\ \citenamefont
  {Gal}(2008)}]{Friedman:2008es}%
  \BibitemOpen
  \bibfield  {author} {\bibinfo {author} {\bibfnamefont {E.}~\bibnamefont
  {Friedman}}\ and\ \bibinfo {author} {\bibfnamefont {A.}~\bibnamefont {Gal}},\
  }\href {\doibase 10.1103/PhysRevD.78.016002} {\bibfield  {journal} {\bibinfo
  {journal} {Phys. Rev. D}\ }\textbf {\bibinfo {volume} {78}},\ \bibinfo
  {pages} {016002} (\bibinfo {year} {2008})},\ \Eprint
  {http://arxiv.org/abs/0803.3696} {arXiv:0803.3696 [hep-ph]} \BibitemShut
  {NoStop}%
\bibitem [{\citenamefont {Chung}\ \emph {et~al.}(2002)\citenamefont {Chung}
  \emph {et~al.}}]{Chung:2002fx}%
  \BibitemOpen
  \bibfield  {author} {\bibinfo {author} {\bibfnamefont {J.}~\bibnamefont
  {Chung}} \emph {et~al.},\ }\href {\doibase 10.1103/PhysRevD.66.032004}
  {\bibfield  {journal} {\bibinfo  {journal} {Phys. Rev. D}\ }\textbf {\bibinfo
  {volume} {66}},\ \bibinfo {pages} {032004} (\bibinfo {year} {2002})},\
  \Eprint {http://arxiv.org/abs/hep-ex/0205093} {arXiv:hep-ex/0205093}
  \BibitemShut {NoStop}%
\bibitem [{\citenamefont {Abe}\ \emph {et~al.}(2015)\citenamefont {Abe} \emph
  {et~al.}}]{Super-Kamiokande:2011idx}%
  \BibitemOpen
  \bibfield  {author} {\bibinfo {author} {\bibfnamefont {K.}~\bibnamefont
  {Abe}} \emph {et~al.} (\bibinfo {collaboration} {Super-Kamiokande}),\ }\href
  {\doibase 10.1103/PhysRevD.91.072006} {\bibfield  {journal} {\bibinfo
  {journal} {Phys. Rev. D}\ }\textbf {\bibinfo {volume} {91}},\ \bibinfo
  {pages} {072006} (\bibinfo {year} {2015})},\ \Eprint
  {http://arxiv.org/abs/1109.4227} {arXiv:1109.4227 [hep-ex]} \BibitemShut
  {NoStop}%
\bibitem [{\citenamefont {Bellerive}\ \emph {et~al.}(2016)\citenamefont
  {Bellerive}, \citenamefont {Klein}, \citenamefont {McDonald}, \citenamefont
  {Noble},\ and\ \citenamefont {Poon}}]{Bellerive:2016byv}%
  \BibitemOpen
  \bibfield  {author} {\bibinfo {author} {\bibfnamefont {A.}~\bibnamefont
  {Bellerive}}, \bibinfo {author} {\bibfnamefont {J.~R.}\ \bibnamefont
  {Klein}}, \bibinfo {author} {\bibfnamefont {A.~B.}\ \bibnamefont {McDonald}},
  \bibinfo {author} {\bibfnamefont {A.~J.}\ \bibnamefont {Noble}}, \ and\
  \bibinfo {author} {\bibfnamefont {A.~W.~P.}\ \bibnamefont {Poon}} (\bibinfo
  {collaboration} {SNO}),\ }\href {\doibase 10.1016/j.nuclphysb.2016.04.035}
  {\bibfield  {journal} {\bibinfo  {journal} {Nucl. Phys. B}\ }\textbf
  {\bibinfo {volume} {908}},\ \bibinfo {pages} {30} (\bibinfo {year} {2016})},\
  \Eprint {http://arxiv.org/abs/1602.02469} {arXiv:1602.02469 [nucl-ex]}
  \BibitemShut {NoStop}%
\bibitem [{\citenamefont {Mohapatra}\ and\ \citenamefont
  {Marshak}(1980)}]{Mohapatra:1980qe}%
  \BibitemOpen
  \bibfield  {author} {\bibinfo {author} {\bibfnamefont {R.~N.}\ \bibnamefont
  {Mohapatra}}\ and\ \bibinfo {author} {\bibfnamefont {R.~E.}\ \bibnamefont
  {Marshak}},\ }\href {\doibase 10.1103/PhysRevLett.44.1316} {\bibfield
  {journal} {\bibinfo  {journal} {Phys. Rev. Lett.}\ }\textbf {\bibinfo
  {volume} {44}},\ \bibinfo {pages} {1316} (\bibinfo {year} {1980})},\ \bibinfo
  {note} {[Erratum: Phys.Rev.Lett. 44, 1643 (1980)]}\BibitemShut {NoStop}%
\bibitem [{\citenamefont {Ozer}(1982)}]{Ozer:1982qh}%
  \BibitemOpen
  \bibfield  {author} {\bibinfo {author} {\bibfnamefont {M.}~\bibnamefont
  {Ozer}},\ }\href {\doibase 10.1103/PhysRevD.26.3159} {\bibfield  {journal}
  {\bibinfo  {journal} {Phys. Rev. D}\ }\textbf {\bibinfo {volume} {26}},\
  \bibinfo {pages} {3159} (\bibinfo {year} {1982})}\BibitemShut {NoStop}%
\bibitem [{\citenamefont {Kuo}\ and\ \citenamefont {Love}(1980)}]{Kuo:1980ew}%
  \BibitemOpen
  \bibfield  {author} {\bibinfo {author} {\bibfnamefont {T.-K.}\ \bibnamefont
  {Kuo}}\ and\ \bibinfo {author} {\bibfnamefont {S.~T.}\ \bibnamefont {Love}},\
  }\href {\doibase 10.1103/PhysRevLett.45.93} {\bibfield  {journal} {\bibinfo
  {journal} {Phys. Rev. Lett.}\ }\textbf {\bibinfo {volume} {45}},\ \bibinfo
  {pages} {93} (\bibinfo {year} {1980})}\BibitemShut {NoStop}%
\bibitem [{\citenamefont {Arnold}\ \emph {et~al.}(2013)\citenamefont {Arnold},
  \citenamefont {Fornal},\ and\ \citenamefont {Wise}}]{Arnold:2012sd}%
  \BibitemOpen
  \bibfield  {author} {\bibinfo {author} {\bibfnamefont {J.~M.}\ \bibnamefont
  {Arnold}}, \bibinfo {author} {\bibfnamefont {B.}~\bibnamefont {Fornal}}, \
  and\ \bibinfo {author} {\bibfnamefont {M.~B.}\ \bibnamefont {Wise}},\ }\href
  {\doibase 10.1103/PhysRevD.87.075004} {\bibfield  {journal} {\bibinfo
  {journal} {Phys. Rev. D}\ }\textbf {\bibinfo {volume} {87}},\ \bibinfo
  {pages} {075004} (\bibinfo {year} {2013})},\ \Eprint
  {http://arxiv.org/abs/1212.4556} {arXiv:1212.4556 [hep-ph]} \BibitemShut
  {NoStop}%
\bibitem [{\citenamefont {Choudhury}\ and\ \citenamefont
  {SenGupta}(2007)}]{Choudhury:2006nj}%
  \BibitemOpen
  \bibfield  {author} {\bibinfo {author} {\bibfnamefont {D.}~\bibnamefont
  {Choudhury}}\ and\ \bibinfo {author} {\bibfnamefont {S.}~\bibnamefont
  {SenGupta}},\ }\href {\doibase 10.1103/PhysRevD.76.064030} {\bibfield
  {journal} {\bibinfo  {journal} {Phys. Rev. D}\ }\textbf {\bibinfo {volume}
  {76}},\ \bibinfo {pages} {064030} (\bibinfo {year} {2007})},\ \Eprint
  {http://arxiv.org/abs/hep-th/0612246} {arXiv:hep-th/0612246} \BibitemShut
  {NoStop}%
\bibitem [{\citenamefont {Arun}\ and\ \citenamefont
  {Choudhury}(2017)}]{Arun:2016csq}%
  \BibitemOpen
  \bibfield  {author} {\bibinfo {author} {\bibfnamefont {M.~T.}\ \bibnamefont
  {Arun}}\ and\ \bibinfo {author} {\bibfnamefont {D.}~\bibnamefont
  {Choudhury}},\ }\href {\doibase 10.1016/j.nuclphysb.2017.08.004} {\bibfield
  {journal} {\bibinfo  {journal} {Nucl. Phys. B}\ }\textbf {\bibinfo {volume}
  {923}},\ \bibinfo {pages} {258} (\bibinfo {year} {2017})},\ \Eprint
  {http://arxiv.org/abs/1606.00642} {arXiv:1606.00642 [hep-th]} \BibitemShut
  {NoStop}%
\bibitem [{\citenamefont {Arun}\ \emph {et~al.}(2017)\citenamefont {Arun},
  \citenamefont {Choudhury},\ and\ \citenamefont {Sachdeva}}]{Arun:2017zap}%
  \BibitemOpen
  \bibfield  {author} {\bibinfo {author} {\bibfnamefont {M.~T.}\ \bibnamefont
  {Arun}}, \bibinfo {author} {\bibfnamefont {D.}~\bibnamefont {Choudhury}}, \
  and\ \bibinfo {author} {\bibfnamefont {D.}~\bibnamefont {Sachdeva}},\ }\href
  {\doibase 10.1088/1475-7516/2017/10/041} {\bibfield  {journal} {\bibinfo
  {journal} {JCAP}\ }\textbf {\bibinfo {volume} {10}},\ \bibinfo {pages} {041}
  (\bibinfo {year} {2017})},\ \Eprint {http://arxiv.org/abs/1703.04985}
  {arXiv:1703.04985 [hep-ph]} \BibitemShut {NoStop}%
\bibitem [{\citenamefont {Tsutsui}\ \emph {et~al.}(2004)\citenamefont {Tsutsui}
  \emph {et~al.}}]{CP-PACS:2004wqk}%
  \BibitemOpen
  \bibfield  {author} {\bibinfo {author} {\bibfnamefont {N.}~\bibnamefont
  {Tsutsui}} \emph {et~al.} (\bibinfo {collaboration} {CP-PACS, JLQCD}),\
  }\href {\doibase 10.1103/PhysRevD.70.111501} {\bibfield  {journal} {\bibinfo
  {journal} {Phys. Rev. D}\ }\textbf {\bibinfo {volume} {70}},\ \bibinfo
  {pages} {111501} (\bibinfo {year} {2004})},\ \Eprint
  {http://arxiv.org/abs/hep-lat/0402026} {arXiv:hep-lat/0402026} \BibitemShut
  {NoStop}%
\bibitem [{\citenamefont {Chakraverty}\ and\ \citenamefont
  {Choudhury}(2001)}]{Chakraverty:2000df}%
  \BibitemOpen
  \bibfield  {author} {\bibinfo {author} {\bibfnamefont {D.}~\bibnamefont
  {Chakraverty}}\ and\ \bibinfo {author} {\bibfnamefont {D.}~\bibnamefont
  {Choudhury}},\ }\href {\doibase 10.1103/PhysRevD.63.112002} {\bibfield
  {journal} {\bibinfo  {journal} {Phys. Rev. D}\ }\textbf {\bibinfo {volume}
  {63}},\ \bibinfo {pages} {112002} (\bibinfo {year} {2001})},\ \Eprint
  {http://arxiv.org/abs/hep-ph/0012309} {arXiv:hep-ph/0012309} \BibitemShut
  {NoStop}%
\bibitem [{\citenamefont {Aoki}\ \emph {et~al.}(2020)\citenamefont {Aoki} \emph
  {et~al.}}]{FlavourLatticeAveragingGroup:2019iem}%
  \BibitemOpen
  \bibfield  {author} {\bibinfo {author} {\bibfnamefont {S.}~\bibnamefont
  {Aoki}} \emph {et~al.} (\bibinfo {collaboration} {Flavour Lattice Averaging
  Group}),\ }\href {\doibase 10.1140/epjc/s10052-019-7354-7} {\bibfield
  {journal} {\bibinfo  {journal} {Eur. Phys. J. C}\ }\textbf {\bibinfo {volume}
  {80}},\ \bibinfo {pages} {113} (\bibinfo {year} {2020})},\ \Eprint
  {http://arxiv.org/abs/1902.08191} {arXiv:1902.08191 [hep-lat]} \BibitemShut
  {NoStop}%
\bibitem [{\citenamefont {Cline}(2006)}]{Cline:2006ts}%
  \BibitemOpen
  \bibfield  {author} {\bibinfo {author} {\bibfnamefont {J.~M.}\ \bibnamefont
  {Cline}},\ }in\ \href@noop {} {\emph {\bibinfo {booktitle} {{Les Houches
  Summer School - Session 86: Particle Physics and Cosmology: The Fabric of
  Spacetime}}}}\ (\bibinfo {year} {2006})\ \Eprint
  {http://arxiv.org/abs/hep-ph/0609145} {arXiv:hep-ph/0609145} \BibitemShut
  {NoStop}%
\end{thebibliography}%

\end{document}